\documentclass[runningheads]{llncs}

\usepackage{float}
\usepackage{color}
\usepackage{graphicx}
\usepackage{hyperref}

\begin{document}
\title{A coupled-mechanisms modelling framework for neurodegeneration}

\author{Tiantian He et al}
\author{Tiantian He\inst{1}\and
Elinor Thompson\inst{1}\and
Anna Schroder\inst{1}\and
Neil P. Oxtoby\inst{1}\and
Ahmed Abdulaal\inst{1}\and
Frederik Barkhof\inst{1,2,3}\and
Daniel C. Alexander\inst{1}}
% index{He, Tiantian} 
% index{Thompson, Elinor} 
% index{Schroder, Anna} 
% index{Oxtoby, Neil P.} 
% index{Abdulaal, Ahmed} 
% index{Barkhof, Frederik} 
% index{Alexander, Daniel C.} 
%
\authorrunning{He et al.}
\institute{
UCL Centre for Medical Image Computing, Department of Computer Science, University College London, London, UK \and
UCL Queen Square Institute of Neurology, University College London, London, UK \and Department of Radiology \& Nuclear Medicine, Amsterdam University Medical Center, Vrije Universiteit, Amsterdam, the Netherlands
\\\email{tiantian.he.20@ucl.ac.uk}
}
\maketitle              % typeset the header of the contribution
\begin{abstract}

Computational models of neurodegeneration aim to emulate the evolving pattern of pathology in the brain during neurodegenerative disease, such as Alzheimer's disease. Previous studies have made specific choices on the mechanisms of pathology production and diffusion, or assume that all the subjects lie on the same disease progression trajectory. However, the complexity and heterogeneity of neurodegenerative pathology suggests that multiple mechanisms may contribute synergistically with complex interactions, meanwhile the degree of contribution of each mechanism may vary among individuals. We thus put forward a coupled-mechanisms modelling framework which non-linearly combines the network-topology-informed pathology appearance with the process of pathology spreading within a dynamic modelling system. We account for the heterogeneity of disease by fitting the model at the individual level, allowing the epicenters and rate of progression to vary among subjects.  We construct a Bayesian model selection framework to account for feature importance and parameter uncertainty. This provides a combination of mechanisms that best explains the observations for each individual from the ADNI dataset. With the obtained distribution of mechanism importance for each subject, we are able to identify subgroups of patients sharing similar combinations of apparent mechanisms.

\keywords{Network spreading model  \and Disease progression \and Bayesian model selection \and Variational Inference .}
\end{abstract}
\footnotetext[1]{Elinor Thompson and Anna Schroder contributed equally to this work as the co-second authors.}

\section{Introduction}

% mechanistic models of ND and why they are important

Computational models of neurodegeneration aim to emulate the underlying physical process of how disease initiates and progresses over the brain from a mechanistic point of view \cite{Garbarino2021InvestigatingBrain,Weickenmeier2018MultiphysicsAtrophy,Iturria-Medina2017MultifactorialDisease}.  
A better understanding of disease mechanisms and inter-individual variability through these models will aid in the development of new treatments and disease prevention strategies.

Two types of components typically contribute to such models: models of pathology appearance - how and where pathology spontaneously appears in different brain regions; and models of pathology spreading - how pathology spreads from region to region. Both components are often linked to brain connectivity so that spontaneous appearance arises according to network topologies, and spreading is facilitated by network edges or brain-connectivity pathways.

Network metrics link spontaneous appearance of pathology to brain connectivity via various mechanistic hypotheses. For instance, Zhou et al. \cite{Zhou2012PredictingConnectome} relate disease patterns with several network topologies: i) centrality \cite{Buckner2005MolecularMemory}, indicating that regions with denser connections are more vulnerable to disease due to heavier nodal stress; ii) segregation \cite{Appel1981ADisease}, the converse of centrality, stating that regions with sparse connections are more susceptible due to lack of trophic sources; iii) shared vulnerability, expounding that connected regions have evenly distributed disease because they have common characteristics such as gene expressions \cite{Iturria-Medina2017MultifactorialDisease}.

%% spreading models 
For the spreading component, dynamical systems models emulate the spatio-temporal propagation process along the brain connectivity \cite{Iturria-Medina2014EpidemicDisorders,Raj2012ADementia,Raj2015NetworkDisease,Iturria-Medina2018MultimodalNeurodegeneration,vogel_spread_2020}. One popular example is the network diffusion model \cite{Raj2012ADementia} (NDM), which assumes that the pathology purely diffuses from epicenters. Weickenmeier et al. \cite{Weickenmeier2018MultiphysicsAtrophy} combine the disease diffusion process with a local production term in a single model, which emulates the full process of how the protein diffuses from epicenters and replicates locally, gradually reaching a plateau. Thus, this model is able to reconstruct the process from disease onset to later stages. 

However, such models make specific choices on the underlying mechanism in the particular physical process. The complexity and heterogeneity of neurodegenerative conditions suggests that multiple processes may contribute and vary among individuals. To avoid making such assumptions, Garbarino et al.\cite{Garbarino2019DifferencesData} use a data-driven, linear combination of several network topological descriptors extracted from the structural connectome to match the disease patterns fitted by a Gaussian Process progression model. They find that the combination, which they refer to as a "mechanistic profile" better matches observed pathology patterns than any single characteristic. However, this considers appearance and spreading as interchangeable rather than interacting mechanisms. Secondly, although \cite{Garbarino2019DifferencesData} does produce individual-level as well as group-level mechanistic profiles, the individual-level profiles assume that all the subjects lie on the same disease progression trajectory. This does not fully capture the heterogeneity (such as different epicenters \cite{Vogel2021FourDisease}, diffusion rate) within groups and underestimates the variability in composition of the mechanistic profile among subjects.

In this work, we introduce an alternative model framework that non-linearly couples the effects of spontaneous appearance and spreading. We construct a Bayesian framework with an appropriate sparsity structure to estimate the mechanistic profile, in a similar way to \cite{Garbarino2019DifferencesData} but including interaction of model components and quantification of uncertainty. We account for the heterogeneity of neurodegenerative diseases in a more complex way than \cite{Garbarino2019DifferencesData}, by allowing factors like epicenters, rates of diffusion and production, and the weights of network metrics to vary among individuals. The resulting mechanistic profiles highlight distinct subgroups of individuals within an Alzheimer's disease (AD) cohort, in which each subgroup has similar combinations of network metrics.

\section{Methodology}

\subsection{Model Definition}
\subsubsection{Baseline Model}  
One model of disease spreading \cite{Weickenmeier2018MultiphysicsAtrophy,Meisl2021InDisease} based on the Fisher-Kolmogorov equation, assumes that the concentration of toxic protein can be emulated by an ordinary differential equation (ODE) system, which involves the combination of two physical processes: 1) the diffusion of toxic proteins along the structural network from an epicentre(s), as described by the NDM \cite{Raj2012ADementia}; 2) its local production and aggregation at each node. The diffusion component includes the graph Laplacian matrix $\mathbf{L}$ calculated from the structural connectome \cite{Raj2012ADementia}, as a substrate for disease diffusion with the rate $k$. The production and aggregation part provides a monotonically increasing trend converging to a plateau level $v$ with a common speed $\alpha$ shared by all regions. The model evolves the pathology concentration $\mathbf{c}$ at time $t$ according to:
\begin{equation}
\label{eq:baseline NDM}
\frac{d \mathbf{c}}{d t}=-k [\mathbf{L} \mathbf{c}(t)]+\alpha \mathbf{c}(t) \odot[v-\mathbf{c}(t)],
\end{equation}
where $\odot$ denotes the element-wise product. It constructs the disease progression process from onset to late stage based on the single mechanism of network proximity. It assumes a constant local production rate across regions, thus the growth of concentration only depends on the level of biomarker propagating from the epicenter along the structural connectome to each node. This does not take into account the synergistic effect of other mechanisms.

\subsubsection{Coupled Model}
Following \cite{Weickenmeier2018MultiphysicsAtrophy}, we retain the network proximity mechanism for the 
diffusion component of our model, but we weight the local production process with the combination of P network metrics  $\mathbf{M} = [\mathbf{m_1},...,\mathbf{m_P}]$. We use $\mathbf{w} = [w_1,...,w_P]^{T}$ to represent the weighting, or extent of contribution of each characteristic. In contrast to the baseline model, by weighting the rate of production with the combination of network metrics, we further incorporate pathology appearance proportional to various brain network topologies to the disease spreading process. The new model expresses this coupling as follows:

\begin{equation}
\label{eq:coupled NDM}
\frac{d \mathbf{c}}{d t}=-k [\mathbf{L} \mathbf{c}(t)]+\alpha  \mathbf{M} \mathbf{w} \odot \mathbf{c}(t) \odot[v-\mathbf{c}(t)].
\end{equation}

\subsubsection{Network Metrics}
\begin{figure}
\centering
\includegraphics[scale=0.29]{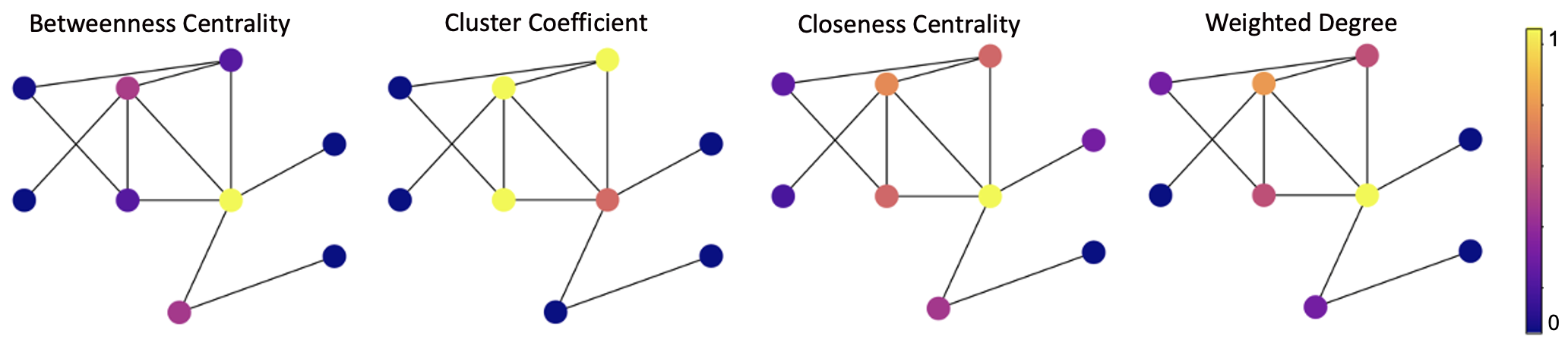}
\caption{\textbf{Visualization of network metrics of the toy brain connectivity.}
$\mathbf{BetweennessCentrality}$ is the fraction of all shortest paths that contain a specific node. $\mathbf{ClusterCoefficient}$ quantifies the extent to which one node is clustered with its neighbours. $\mathbf{ClosenessCentrality}$ is proportional to the reciprocal of the sum of the shortest path length between the node and others. $\mathbf{WeightedDegree}$ is the sum of the weights of the connections. Calculation of metrics has been done using the Brain Connectivity Toolbox \cite{rubinov2010complex}. Metrics $\mathbf{m}_1$ to $\mathbf{m}_5$ have been normalized between [0,1] to maintain the same scale. Visualization is done using NetworkX (https://networkx.org/).} 
\label{fig:NetworkX}
\end{figure}

The network metrics considered by the model are listed in Table \ref{sigmoid_metrics} and a visualization is shown in Figure \ref{fig:NetworkX}. $\mathbf{StructureC}$, $\mathbf{InvGeodist}$, $\mathbf{FunctionC}$ represent the structural connectome, the matrix of the inverse of geodesic distance along the cortical surface and the functional connectome.
 
\begin{table}
\centering
\caption{Network metrics used as the interaction component with the spreading model}\label{sigmoid_metrics}
\begin{tabular}{|l|l|l|l|}
\hline
Metrics &  formula & weights & mechanism\\
\hline
$\mathbf{m}_1$ &  $\mathbf{1-BetweennessCentrality(StructureC)}$ &$w_1$ & Structural Segregation\\
$\mathbf{m}_2$ & $\mathbf{1-ClusterCoefficient(StructureC)}$ & $w_2$ & Structural Dispersion\\
$\mathbf{m}_3$ & $\mathbf{ClosenessCentrality(StructureC)}$ & $w_3$ & Structural Centrality\\
$\mathbf{m}_4$ & $\mathbf{WeightedDegree(InvGeodist)}$ & $w_4$ & Geodesic proximity\\
$\mathbf{m}_5$ & $\mathbf{BetweennessCentrality(FunctionC)}$ & $w_5$ & Functional centrality\\
$\mathbf{m}_6$ & $\mathbf{1}$ & $w_6$ & Even distribution\\
\hline
\end{tabular}
\end{table}

\subsection{Bayesian Framework}
In order to quantify the uncertainty of the estimation, we construct a Bayesian inference framework for our dynamic system, thus we are able to obtain distributions of parameters rather than deterministic values.
\subsubsection{Parameter distributions}
In this work we focus on modelling the dynamics of tau protein, which is widely hypothesized to be a key causative agent in AD. Its concentration can be measured in vivo by positron emission tomography (PET). We assume for subject $i$ at $j$th scan time $t_{ij}$, the measurement of tau concentration $\hat{\mathbf{c}}(t_{ij})$ follows a normal distribution 
\begin{equation}
p(\hat{\mathbf{c}}(t_{ij}) \mid k_i,\alpha_i) \sim Normal\left(\mathbf{c}(t_{ij}, k_i,\alpha_i), \sigma^{2}\right),    
\end{equation}
where the mean is the model prediction with best-fit parameters and the error is quantified by the standard deviation $\sigma$. 
The time gap $\delta_{ij}$ (in years) between the baseline scan and the $j$th follow-up scan is available in the dataset. However the time from the disease onset to the baseline scan is unknown. Thus we need to estimate such time $t^{onset}_i$ such that $t_{ij} = t^{onset}_i + \delta_{ij}$. 
This time parameterization enforces the relevant locations among all scans fixed by given $\delta_{ij}$.

Descriptions of the key model parameters are displayed in Table \ref{param_baseline}. We enforce the rate of spreading and production to be positive by selecting a Half-Normal prior. The hyper-parameters of the rates are decided according to the research findings that the annual change rate of tau-PET signal is quite slight \cite{jack2020predicting,smith2020accumulation}. Explorations of the proper choice of the hyper-parameters has also been done by simulating different parameter levels and comparing the generated trajectories to the measured data distribution. 

\begin{table}
\centering
\caption{Model parameters and their prior distribution}\label{param_baseline}
\begin{tabular}{|l|l|l|l|}
\hline
Baseline Model & Coupled model & Interpretation & Prior distribution \\
\hline
$k$ &  $k$ & Spreading rate& HalfNormal(std = 0.1) \\
$\alpha$ & $\alpha$ &Local production rate &HalfNormal(std = 0.1)\\
$t_{onset}$ &  $t_{onset}$ & Pseudo onset time & Uniform(10,100) \\
$\sigma$ &$\sigma$ &Overall uncertainty& HalfNormal(std = 1)\\
 &$\mathbf{w}$& Feature weights& Dirichlet($\beta_1 \ldots \beta_P$)\\
\hline
\end{tabular}
\end{table}

\subsubsection{Feature selection with sparsity}

We account for the feature importance by estimating the weights of network metrics 
$$
\mathbf{w} = [w_1,...,w_P]^{T} \sim Dirichlet(\beta_1, \ldots, \beta_P).
$$
We seek the minimal set of network components that explain the data to define the mechanistic profile. 
Thus, we apply sparsity to the weight in a Bayesian way by introducing the Horseshoe prior \cite{Carvalho2009HandlingHorseshoe} to the hyper-parameters of the Dirichlet distribution: 
$$
\beta_l \mid \lambda_l, \tau \sim HalfNormal\left(0, \lambda_l^2 \tau^2 \right), l = 1 \ldots P.
$$
This horseshoe structure includes a global shrinkage parameter $\tau$ and local shrinkage parameters $\lambda_l$, each following a half Cauchy distribution:
$$
\lambda_l, \tau \sim HalfCauchy(0,1).
$$
The flat tail from Cauchy Distribution allows the features with strong contribution to remain with a heavy tail of the density, while the sharp rise in density near 0 shrinks the weight of the features with weak signal.

\subsection{Variational Inference}
Suppose $\mathbf{x}$, $\mathbf{z}$ and $\theta$ represent the collections of observations, hidden variables and parameters respectively. 
Due to the complexity of the model structure, the posterior $p_{\theta}(\mathbf{z} \mid \mathbf{x})$ we are interested in is often intractable and hard to obtain analytically. Thus we use the variational distribution $q_\phi(\mathbf{z})$ with $\phi$ as the variational parameters to approximate the posterior. 
The evidence lower bound,
$
\mathrm{ELBO} \equiv \mathrm{E}_{q_\phi(\mathbf{z})}\left[\log p_\theta(\mathbf{x}, \mathbf{z})-\log q_\phi(\mathbf{z})\right]
$,
can be used to approach the log likelihood, since the gap between them is the Kullback-Leibler divergence between the variational distribution and the posterior, which is larger or equal to 0:

\begin{equation}
    \log p_\theta(\mathbf{x})-\mathrm{ELBO}=\mathrm{KL}\left(\mathbf{q}_\phi(\mathbf{z}) \| \mathrm{p}_\theta(\mathbf{z} \mid \mathbf{x})\right) \geq 0.
\end{equation}
Thus the objective of the optimization is to maximize the ELBO. We use a normal distribution with a diagonal covariance matrix as the variational distribution to sample the hidden variables in the latent space, and then use proper parameter transformation to obtain the constrained hidden variables. The process is accomplished with the use of Pyro \cite{Bingham2019Pyro:Programming}, a probabilistic programming framework.

\section{Experiments and results}

\subsection{Data processing}
\subsubsection{Brain Networks}
Three types of connectivity were used to extract features for our coupled models: 1) the structural connectome, which contains the number of white matter fibre trajectories; 2) the matrix of the geodesic distance along the cortical surface; 3) the functional connectome, which reflects the synchrony of neural functioning among regions. The structural connectome is an average of 18 young and healthy subjects’ connectomes from the Human Connectome Project \cite{van_essen_wu-minn_2013}. We generated streamlines using probabilistic, anatomically constrained tractography, processed using MRtrix3 \cite{tournier_mrtrix3_2019}, and filtered the streamlines using the SIFT algorithm \cite{smith_sift2_2015}. The geodesic distance matrix and the functional connectome are obtained from the Microstructure-Informed Connectomics Database \cite{Royer2022AnNeuroscience}, and is an average of 50 healthy subjects’ matrices. We define the brain regions according to the Desikan-Killiany Atlas \cite{desikan_automated_2006}. 

\subsubsection{Tau-PET data}
We model the dynamics of tau protein measured by PET scans. We use the tau-PET standardized uptake value ratios (SUVRs) downloaded from the Alzheimer’s Disease Neuroimaging Initiative (ADNI) (adni.loni.usc.edu) \cite{landau_flortaucipir_2021}. We exclude subcortical regions, which are impacted by off-target binding of the radiotracer \cite{groot_tau_2022}. Two-component Gaussian mixture modelling is applied to the SUVR signals of all subjects. We treat the distribution with the lower mean as the distribution of negative signals, and define the mean plus one standard deviation of this distribution as the threshold for tau-positivity.

\subsubsection{Selection of subjects}
We include N = 110 subjects with at least two tau-PET scans, amyloid beta positive status and at least one region with positive tau signal, including healthy, cognitively impaired and AD subjects, since we aim to focus on the people with a potential to accumulate abnormal pathological tangles. We normalise the data of all the subjects (i = 1,..,N) between 0 and 1 by $(tau_{i} - tau_{min}))/(tau_{max}-tau_{min})) $
where $tau_{min}$ and $tau_{max}$ are calculated across all the subjects and regions, thus the differences in data scales among subjects and regions are maintained. 

\subsubsection{Setting of epicentres}
For initialization, we assume pathology starts from candidate epicentres, to simulate the full process of disease progression from very early stages, i.e. prior to the baseline scan. We rank 34 pairs of bilateral cortex regions according to the total number of subjects that have positive tau signals, and pick the top eight pairs of regions as the candidate epicentres where the propagation of pathology is likely to start: inferior temporal cortex, banks of the superior temporal sulcus, fusiform gyrus, lateral orbitofrontal cortex, middle temporal gyrus, entorhinal cortex, parahippocampal gyrus and temporal pole. The four epicentres identified by Vogel et al.\cite{Vogel2021FourDisease} are all included.  
\subsection{Results}
We fix the initial tau level at each candidate epicenter and the end level of the plateau of all the subjects to be 0.01 and 1.5 and fit each subject using the baseline model and our coupled model respectively. For evaluation we use Pearson R correlation between the measured and model fitted values. 

\begin{figure}[!t]
\centering
\includegraphics[scale=0.16]{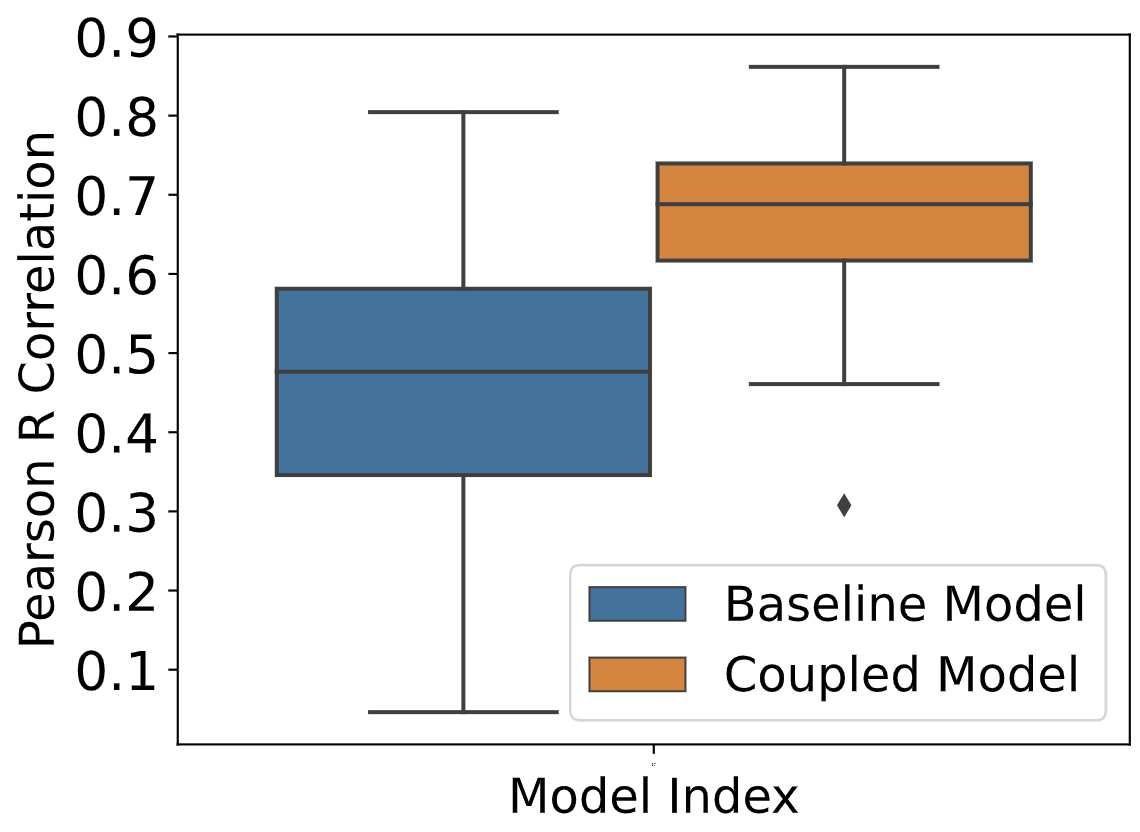}
\caption{$\textbf{Comparison of the overall fitting among all subjects}$. This boxplot visualizes the distribution of Pearson R correlations (between model fitting and the measured value) of all subjects fitted by the baseline and our coupled model respectively.} \label{Model comparison}
\end{figure}

\begin{figure}
\centering
\includegraphics[scale = 0.21]{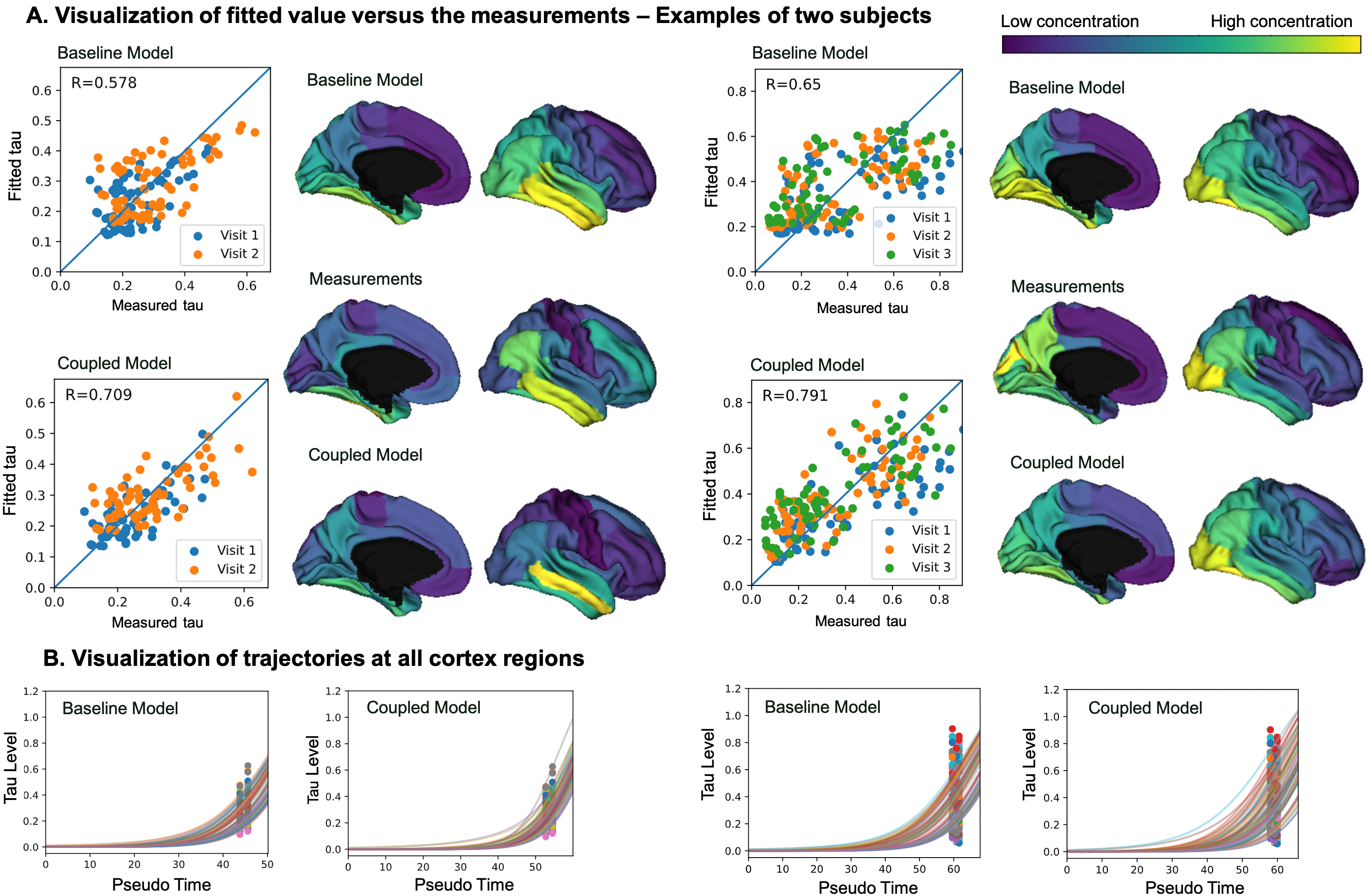}

\caption{\textbf{Model comparison at individual level.}
We display examples from two subjects by comparing their fitting from the baseline model and the coupled model. A: scatter plots which show the model-fitted signals versus the measured signals and the visualization of signals on the brain. Each point in the scatter plots represents the signal in one region. B: visualizations of the individual-level trajectory. Each line represents the modelled trajectory for a region while each point represents the actual signal in that region at each visit. Improvement can be observed by the coupled model.} \label{all trajectory}
\end{figure}

Figure \ref{Model comparison} compares the distribution of R correlations which reflect the performance of the baseline model and our coupled model on each individual. According to the one-side t test, the R correlations from the coupled model are significantly larger than those from the baseline model (p-value=3.5732e-22). Especially, fitting of subjects with particularly low performance in the baseline model are noticeably improved.

\begin{figure}
\centering
\includegraphics[scale=0.18]{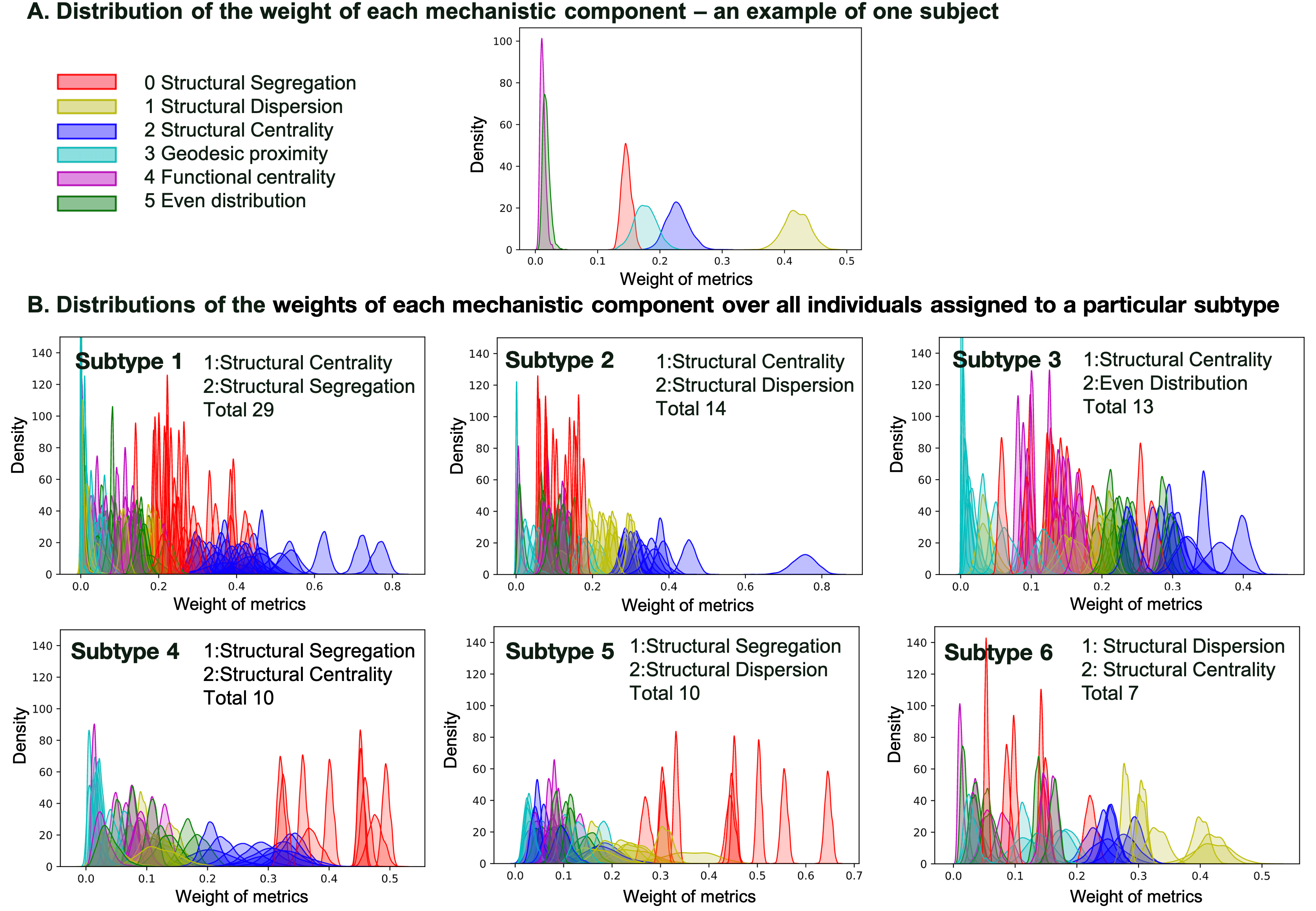}
\caption{Distributions of the weights for each mechanistic component of an individual (A) and over all individuals belonging to each particular subtype obtained based on the top two important features (B).} \label{subtypes}
\end{figure}

Figure \ref{all trajectory} visualizes the model-fitted tau signals versus measured tau signals in all 68 cortex regions. The distribution of tau fitted by our coupled model is closer to the measured pattern.
Finally, we subtype the individuals according to the top two most dominant mechanisms of pathology appearance interacting with the network spreading model. Specifically, we encode each subject with a vector containing the rank of each metric according to the obtained weights, and assign the subjects having the same rank of the top two metrics to the same group. We consider a group containing at least 6 people (5\% of all) as one subtype. As a result, 83 out of 110 subjects have been assigned to six subtypes. Figure \ref{subtypes}A displays the feature importance distribution for an individual, while Figure \ref{subtypes}B places the feature distributions of the subjects belonging to the same subtype within each of the six plots. It can be observed that structural centrality appears most frequently as a dominant feature, followed by structural segregation. 

\section{Conclusions}
We introduce a new Bayesian modelling framework that couples pathology appearance and spreading, by embedding
the mechanistic profiles that consist of combinations of network metrics into the dynamic system of disease spreading. This improves the fitting of the observed pathology pattern, and provides a potential way to subtype subjects according to mechanistic profiles.
For future work, we will validate the cohort-level mechanistic profiles derived from the identified subtypes using external datasets, and also verify the subtypes using other algorithms such as the SuStaIn \cite{young2018uncovering}. Furthermore, we will incorporate uncertainty from connectomes. We will also perform further comparisons with other state-of-the-art models, such as the topological profiles by \cite{Garbarino2019DifferencesData}, which is currently hard to compare directly due to various differences in the model design.

\section{Acknowledgment}
TH, AS and AA are supported by the EPSRC funded UCL Centre for Doctoral Training in Intelligent, Integrated Imaging in Healthcare[EP/S021930/1]; TH, ET and DCA are supported by the Wellcome Trust(221915); DCA and FB are supported by the NIHR Biomedical Research Centre at UCLH and UCL; NPO acknowledges funding from a UKRI Future Leaders Fellowship(MR/S03546X/1). 
Data used in this article were obtained from the Alzheimer’s Disease Neuroimaging Initiative (ADNI) database (adni.loni.usc.edu). As such, the investigators within the ADNI contributed to the design and implementation of ADNI and/or provided data but did not participate in analysis or writing of this report. A complete listing of ADNI investigators can be found at: \url{http://adni.loni.usc.edu/wp-content/uploads/how_to_apply/ADNI_Acknowledgement_List.pdf}.

\bibliographystyle{splncs04}
%\bibliography{miccai_references.bib}
\bibliography{paper}

\newpage
\appendix
\title{Supplementary: A coupled-mechanisms modelling framework for neurodegeneration}
\author{Tiantian He et al}
\institute{
}
\maketitle              % typeset the header of the 

\section{A single topology metric vs our combination of metrics}
\begin{figure}[H]
\centering
\includegraphics[scale = 0.24]{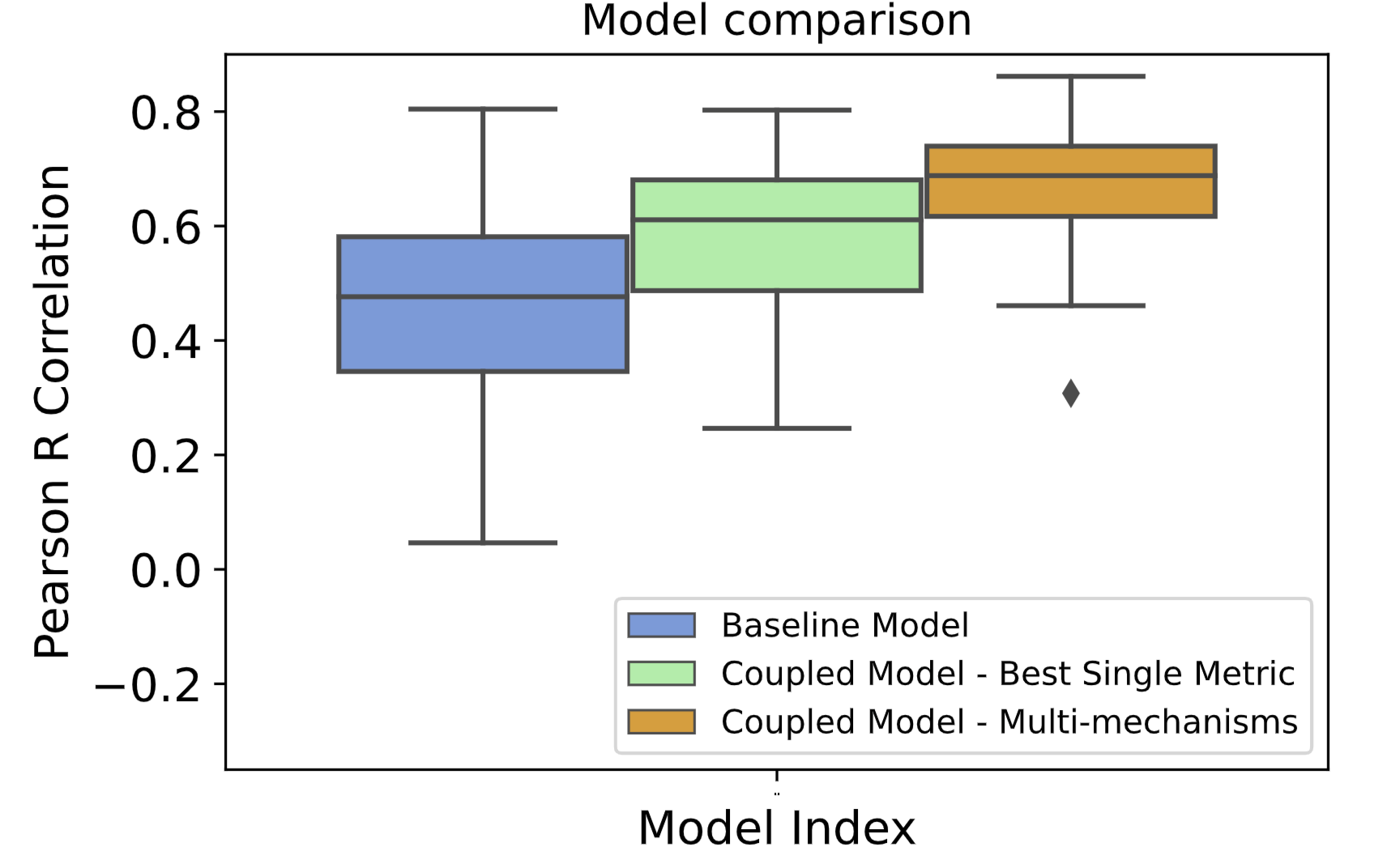}
\caption{\textbf{Weighting the production term with a single metric vs our combination of metrics} We display the model performance (using R correlation of the modelled and observed data) for each subject using the model with its corresponding best single metric and compare it with our multi-mechanisms model. It can be observed that using a single network metric to weight the production term works better than the baseline model, but not as well as the combination of multiple metrics. This suggests that multiple mechanisms of pathology appearance may work synergistically.} 
\label{single_metric}
\end{figure}

\section{Stability test: repeated experiments}

\begin{table}
\label{table:metrics list}
\centering
\caption{\textbf{Summary of model performance across all subjects in three runs} To test the stability of the result, we repeated the experiments three times with different random seeds. Adam is used as the optimizer with a learning rate of 0.01. The table summarizes the overall performance of the baseline and our coupled model across three runs by displaying the average R correlation (fitting vs measurement) across all subjects and all visits. }\label{repeated_table}
\begin{tabular}{|c|c|}
\hline
Model &  Average R correlation in three runs\\
\hline
Baseline model & [0.466, 0.467, 0.464] \\
Coupled model & [0.669, 0.654, 0.670] \\
\hline
\end{tabular}
\end{table}

\begin{figure}[H]
\centering
\includegraphics[scale = 0.27]{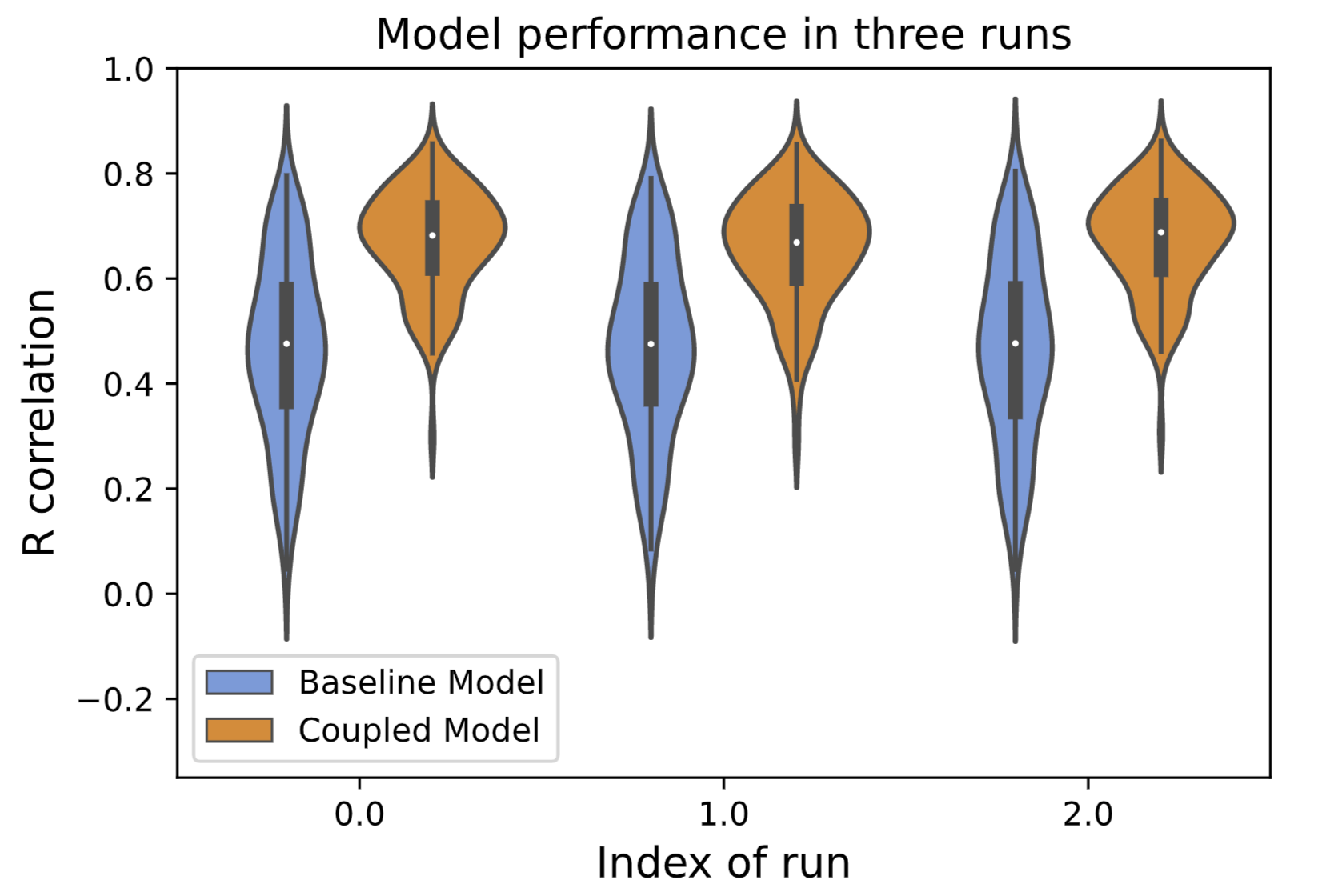}
\caption{\textbf{Repeated experiments across three runs} The results for these three runs are displayed by violin plots, which compare the distribution of R correlations between the model prediction and the measured pathology for each individual. The performance is stable regardless of the randomness.} 
\label{repeated_result}
\end{figure}

\section{Weighting the production term with randomised vectors}
\begin{figure}[H]
\centering
\includegraphics[scale = 0.27]{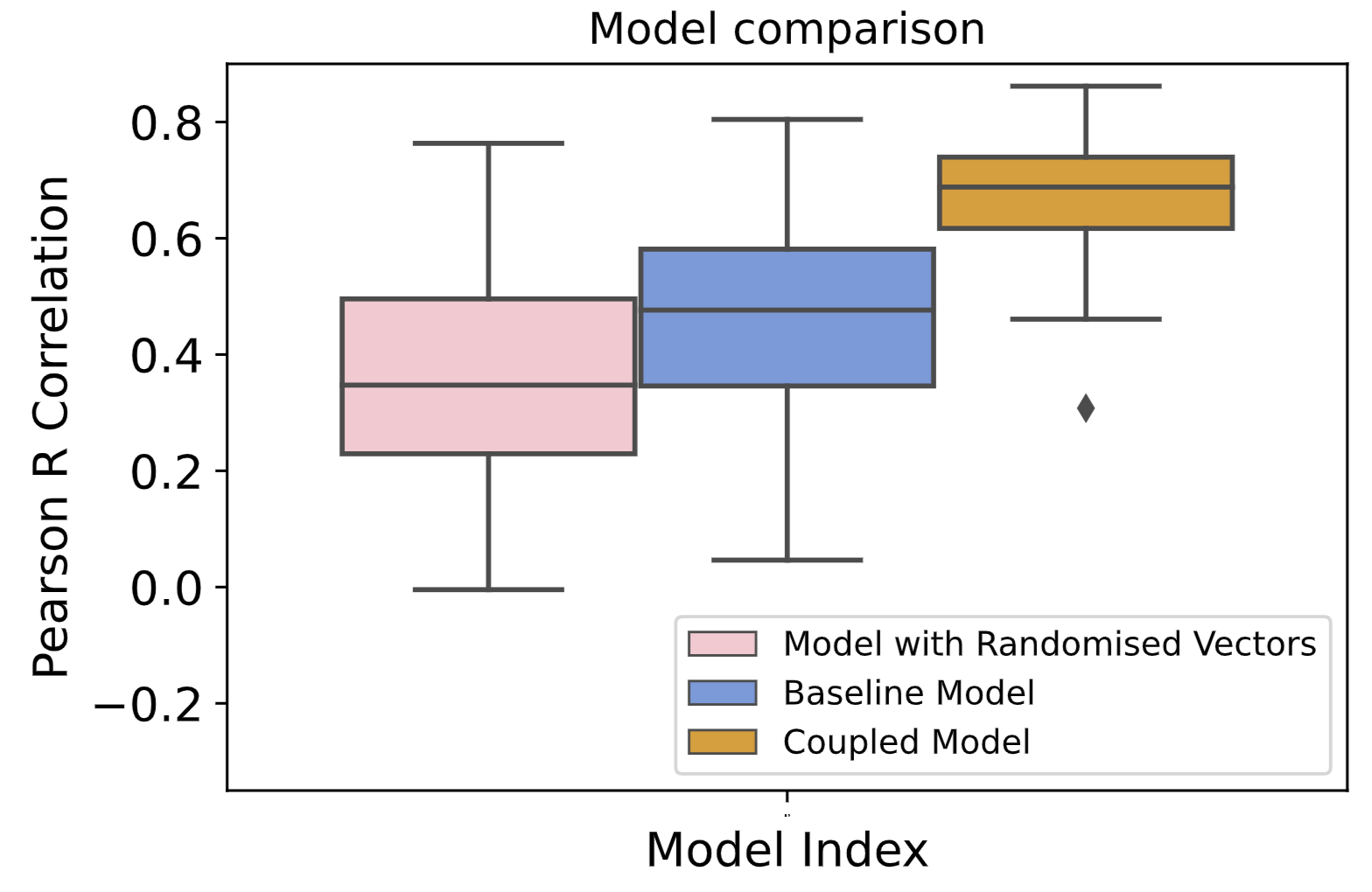}
\caption{\textbf{Using randomised vectors with no biological meaning} We replace the six appearance metrics in our coupled model with six randomised vectors generated from the standard uniform distribution to verify that the model truly benefits from the additional information of the biologically-relevant network metrics. We repeat optimization for all the parameters and weights. Similarly, we use boxplots to display R correlations between the fitting and measurements of all subjects. The use of randomised vectors results in poor model performance, compared to the use of network metrics in the coupled model. This highlights the importance of the network metrics.} 
\label{randomed_metric}
\end{figure}

\end{document}